\documentstyle[sprocl]{article}

\newcommand{\veps}{\mbox{\boldmath $\epsilon$ \unboldmath}}
\newcommand{\valf}{\mbox{\boldmath $\alpha$ \unboldmath}}
\newcommand{\vtau}{\mbox{\boldmath $\tau$ \unboldmath}}  
\newcommand{\vpi}{\mbox{\boldmath $\pi$ \unboldmath}}

\begin{document}

\title{ Role of nucleon resonance excitation \\
in $\phi$ meson photoproduction }

\author{ Q. Zhao$^a$, J.-P. Didelez$^a$, M. Guidal$^a$, B. Saghai$^b$ }
  
\address{$^a$ Institut de Physique Nucleaire, F-91406 Orsay Cedex, France\\
$^b$ Service de Physique Nucleaire, DAPNIA-CEA/Saclay, \\
F-91191 Gif-sur-Yvette Cedex, France }

\maketitle\abstracts{
The resonance effects are investigated in the $\phi$ meson 
photoproduction near threshold 
through a quark model approach with an effective 
Lagrangian. The diffractive contribution is consistently 
estimated by the {\it t}-channel Pomeron exchange. 
Another non-diffractive process, {\it t}-channel $\pi^0$ exchange 
is also included. 
The numerical result shows 
that the Pomeron exchange plays dominant role in the $\phi$ 
meson photoproduction, while the cross sections of the 
non-diffractive processes, i.e., {\it s}- and {\it u}-channel excitations, and 
{\it t}-channel $\pi^0$ exchange, are quite small.
In the polarization observables, we find that large asymmetries
are produced in the backward direction by the interferences from
the {\it s}- and {\it u}-channel resonances,
while in the forward direction, 
only very small asymmetries are generated. 
Meanwhile, we find that the effects from the $\pi^0$ exchange are 
generally negligible. }

%\newpage

\section{ Introduction}

A well established feature in the $\phi$ meson photoproduction
at low momentum transfer and high energies is that the diffractive
scattering governs the reaction mechanism~\cite{Bauer}. 
With the advent of the new high intensity beam
facilities, like JLAB, ELSA, GRAAL, SPring-8, the study of this field in 
both low energy ($E_{\gamma} \sim 2$ GeV) and/or
high momentum transfer ($-t \geq 1$ (GeV/c)$^2$) becomes possible. 
In these regimes, deviations, i.e. non-diffractive reactions,
from the pure diffractive phenomena are expected, which should show up, 
especially in polarization observables. 
In this work, our purpose is to investigate the role played by the 
non-diffractive resonances in the $\phi$ meson photoproduction
near threshold. 

To reach this purpose, the first step must be to describe the diffractive 
phenomena  in consistence with the experimental dat, i.e. 
to reproduce the  energy-independent feature of the cross sections 
in $\gamma p\to \phi p$ at high energy regions, based on which 
the non-diffractive cross sections can be reasonably estimated.
In this work, we introduce a {\it t}-channel Pomeron exchange 
model~\cite{Donnachie,Laget,Haakman,lee} 
to account for the diffractive phenomena in $\gamma p\to \phi p$ 
at high energies. In this model, 
the Pomeron is treated as a $C=+1$ isoscalar photon.

At present time, there is no systematic investigation of the
role played by the resonances in the $\phi$ meson photoproduction. 
In fact, at hadronic level, 
the unknown $\phi NN^*$ couplings have been the barrier
to go further to include the resonances
 since one has to introduce 
at least one parameter for each $\phi NN^*$ coupling vertex, 
therefore, a large number of parameters will appear in the theory. 
On this point, 
the quark model approach shows great advantages: in the 
exact $SU(6)\otimes O(3)$ symmetry
limit, the quark-vector-meson interactions can be described by the 
effective Lagrangian with only two parameters~\cite{plb98,prc98}.

The contribution from $\pi^0$ exchange in $\gamma p\to \phi p$
is quite negligible in 
comparison with not only the Pomeron exchange in the small $|t|$ region, 
but also the resonance contributions in the large $|t|$ region 
though in some polarization observables the interference
between the {\it natural parity} Pomeron exchange 
and the {\it unnatural parity} 
$\pi^0$ exchange produces sizable asymmetries.

We do not include the amplitude for the $\eta$ exchange in this work,  
since although the $\phi$ meson has 
larger decay branching ratio for $\phi\to \eta\gamma$ than 
for $\phi\to \pi^0\gamma$, 
a recent analysis~\cite{LS-1} of the $\eta$ photoproduction
shows that  $g_{\eta NN}$ is smaller than $g_{\pi NN}$ by roughly
a factor of 7, 
which leads to negligible effects due to the $\eta$ exchange  
in the $\phi$ production. 

The rest part of this work is arranged as follows.
In section 2, a brief introduction is given to the formalism
of the model. In section 3, the numerical results for the cross sections
and the polarization observables are presented. 
Discussion and conclusion are given in section 4. 

\section{ The model}
%
% ----- Effective Lagrangian and transition amplitudes -----
%
\subsection{ {\it s}- and {\it u}-channel resonance excitations}

We first introduce the quark model approach with effective Lagrangian
to the {\it s}- and {\it u}-channel resonance contributions 
in $\gamma p\to \phi p$.
The quark-vector-meson coupling is described by the effective Lagrangian:
\begin{equation} \label{3.0}  
L_{eff}= \overline{\psi}(a\gamma_\mu +  
\frac{ib\sigma_{\mu\nu}q^\nu}{2m_q}) \phi^\mu_m \psi,
\end{equation}  
where the quark field $\psi$ can be $u$, $d$, or $s$ for the light-quark
baryon system. $\phi^\mu_m $ represents the vector meson field for  
the light vector mesons ($\omega$, $\rho$, $K^*$ and $\phi$).   
The 3-quark baryon system is described by the NRCQM in the 
$SU(6)\otimes O(3)$ symmetry limit in this calculation.
The vector meson is treated as an elementary point-like particle 
which couples to the constituent quark through the effective interaction.
$a$ and $b$ are the two parameters introduced in the {\it s}- and 
{\it u}-channel.

At tree level, the transition amplitude from the effective Lagrangian 
can be expressed as the contributions from the {\it s}-, {\it u}- and 
{\it t}-channel processes:
\begin{equation}  
M_{fi}=M^s_{fi}+M^u_{fi}+M^t_{fi} \ .  
\label{3.1}  
\end{equation}  

In $\gamma p\to \phi p$, 
$M^t_{fi}$ vanishes since it is protortional to the charge of the
final state $\phi$ meson. 
Then, with the intermediate states introduced, 
the {\it s}- and {\it u}-channel amplitudes can be written as:
\begin{eqnarray}  
M^{s+u}_{fi}&=&i\omega_\gamma\sum_{j}\langle N_f|H_m|N_j\rangle\langle   
N_j|\frac{1}{E_i+\omega_\gamma-E_j}h_e|N_i\rangle\nonumber\\  
&&+i\omega_\gamma\sum_{j} \langle N_f|h_e\frac{1}{E_i-\omega_\phi-E_j}  
|N_j\rangle\langle N_j|H_m|N_i\rangle, 
\label{3.2}  
\end{eqnarray} 
with 
$H_m=-\overline{\psi}(a\gamma_\mu +\frac{ib\sigma_{\mu\nu}  
q^\nu}{2m_q}) \phi^\mu_m \psi$ for the quark-meson coupling vertex, and
\begin{eqnarray}  
h_e=\sum_{l}e_l{{\bf r}_l\cdot{\veps \hskip -0.16 cm }_\gamma}(1-\valf\cdot  
{\hat{\bf k}})e^{i{\bf k\cdot r}_l},~
{\bf{\hat k}}=\frac{\bf k}{\omega_\gamma},
\end{eqnarray} 
where ${\bf k}$ and $\omega_\gamma$ are the three-momentum and energy 
of the incident photon, respectively.
$|N_j\rangle$ represents the complete set of the intermediate 
states. in the NRCQM, those low-lying states ($n\leq 2$)
have been successfully related to the resonances which can be 
taken into account explicitly in the formula. For those higher 
excited states, they can be treated degenerate to the main quantum 
number $n$ in the harmonic oscillator basis. Detailed description 
of this approach can be found in Ref.~\cite{plb98,prc98}. 
In Table 1, resonances in the {\it s}-channel with their 
assignments in the $SU(6)\otimes O(3)$ NRCQM 
symmetry limit are listed.

\begin{table}[htb]
\caption{ Resonances in the {\it s}-channel with their 
assignments in the $SU(6)\otimes O(3)$ NRCQM 
symmetry limit (PDG1998). $M_R$ and $\Gamma_T$ represent
the mass and total width of a resonance, respectively. } 
\protect\label{tab:reso}
\begin{center}
\begin{tabular}{cccc}
\hline
Resonances & $SU(6)\otimes O(3)$ & $M_R$ (MeV) & $\Gamma_T$ (MeV) \\
\hline
$S_{11}(1535)$ & $N(^2P_M)_{{\frac 12}^-}$ & 1535 & 150\\
$D_{13}(1520)$ & $N(^2P_M)_{{\frac 32}^-}$ & 1520 & 120\\
$P_{13}(1720)$ & $N(^2D_S)_{{\frac 32}^+}$ & 1720 & 150\\
$F_{15}(1680)$ & $N(^2D_S)_{{\frac 52}^+}$ & 1680 & 130\\
$P_{11}(1440)$ & $N(^2S^\prime_S)_{{\frac 12}^+}$ & 1440 & 350\\
$P_{11}(1710)$ & $N(^2S_M)_{{\frac 12}^+}$ & 1710 & 100\\
$P_{13}(1900)$ & $N(^2D_M)_{{\frac 32}^+}$ & 1900 & 400\\
$F_{15}(2000)$ & $N(^2D_M)_{{\frac 52}^+}$ & 2000 & 450\\
\hline
\end{tabular}
\end{center}
\end{table}

%
% ----- t-channel diffractive Pomeron exchange -----
%
\subsection{ t-channel diffractive Pomeron exchange }

As one has known that the {\it t}-channel diffractive process
plays dominant role in $\gamma p\to \phi p$. The main feature 
of the diffractive contribution is that the total cross sections
exhibit almost energy-independent which is accounted for by the 
{\it t}-channel Pomeron exchange model by 
Donnachie and Landshoff based on the Regge phenomenology. 
In this model, the Pomeron  
mediates the long range interaction between 
two confined quarks, and behaves rather like a $C=+1$ isoscalar photon. 

We summarize the vertices and form factors as follows:
\begin{itemize}
\item Pomeron-nucleon coupling: 
\begin{equation}
F_{\mu}(t)= 3\beta_0\gamma_{\mu}f(t), \  
f(t)= \frac{(4M^2_N-2.8t)}{(4M^2_N-t)(1-t/0.7)^2} \ .
\end{equation}
where $\beta_0$ is the coupling of the Pomeron to one light constituent quark. 
$f(t)$ is the isoscalar nucleon electromagnetic form factor. The factor 3 
comes from the ``quark-counting rule".

\item Quark-$\phi$-meson coupling: 
\begin{equation}
V_\nu(p-\frac 12 q, p+\frac 12 q)=f_\phi M_\phi\gamma_\nu \ , \
\Gamma_{\phi\to e^+e^-}=\frac{8\pi \alpha^2_e e^2_Q}{3}
(\frac{f^2_\phi}{M_\phi}) \ .
\end{equation}
where $f_\phi$ is the decay constant of the $\phi$ meson in $\phi\to e^+ e^-$, 
which is determined by the decay width $\Gamma_{\phi\to e^+e^-}$. 

\item Form factor for the Pomeron-off-shell-quark vertex:
\begin{equation}
\mu^2_0/(\mu^2_0+p^2) \ .
\end{equation}
where $\mu_0=1.2$ GeV is the cut-off energy scale for the 
Pomeron-off-shell-quark vertex, and $p$ is the four-momentum of the quark.

\item Pomeron trajectory:
\begin{equation}
\alpha(t)=1+\epsilon+\alpha^\prime t \ , \ \alpha^\prime=0.25 GeV^{-2} \ .
\end{equation}

\end{itemize}

%
% ----- t-channel pion exchange ----
%
\subsection{ {\it t}-channel $\pi^0$ exchange }

The $\pi^0$ exchange is introduced with
the Lagrangian for the $\pi NN$ coupling 
and $\phi\pi\gamma$ coupling as the followings:
\begin{eqnarray}\label{3}
L_{\pi NN}=-i g_{\pi NN}\overline\psi \gamma_5(\vtau\cdot\vpi)\psi \ .
\end{eqnarray}
and 
\begin{eqnarray}\label{4}
L_{\phi \pi^0 \gamma}=e_N\frac{ g_{\phi\pi\gamma} }{M_\phi}
\epsilon_{\alpha\beta\gamma\delta}\partial^\alpha A^\beta
\partial^\gamma\phi^\delta\pi^0 \ .
\end{eqnarray}
Then, the amplitude for the $\pi^0$ exchange can be derived in the NRCQM.
The commonly used couplings, 
${g^2_{\pi NN}}/{4\pi}= 14,
~g^2_{\phi\pi\gamma}=0.143  $, are adopted.

%\begin{center} \underline{\bf Parameters in the model}
%\end{center}
In summary, the parameters appearing in this model are the followings:

\begin{itemize}
\item For the {\it t}-channel Pomeron exchange terms, 
$$\beta_0=1.27\mbox{GeV}^{-1}$$
is determined by data at $E_\gamma=6.45$ GeV.

\item For the {\it s}- and {\it u}-channel contribution, 
numerical fitting of the sparse data (Ref.~\cite{phidata}) gives: 
$$a = -0.035 \pm 0.166~;~ b^\prime\equiv b-a = -0.338 \pm 0.075 \ . $$
We hence fix: $|a|=0.15 \ ,~|b^\prime | = 0.3 $ with the reasons: 
i) The extreme value $|a|=0.15$ will best show the sensitivity 
of observables to $a$; 
ii) No data available for large angles which 
might change the phase of $b^\prime$. 

\item For the $\pi^0$ exchange terms, $\alpha_\pi=300$ MeV is adopted 
for the quark potential in which an exponential factor 
$e^{-\frac{({\bf q-k})^2}{6\alpha^2_\pi}}$ plays a role as a form 
factor for the $\pi NN$ and $\phi\pi\gamma$ vertices. 
This factor comes out naturally in the harmonic oscillator basis
where the nucleon is treated as a non-point-like 3-quark system.

\end{itemize}
%
% ----- Cross section and the polarization observables -----
%
\section{ Cross sections and the polarization observables }
In this work, we limit the discussion to the low energy region 
near the $\phi$ meson production threshold, where the effects
from nucleon resonances are still expected to play a role. 
In the $SU(6)\otimes O(3)$ NRCQM symmetry limit, 
the differential cross section, four single polarization asymmetry 
observables, and the beam-target double polarization asymmetry
are investigated. 

In Fig. 1, the differential cross section for $\gamma p\to \phi p$ 
at $E_\gamma=2.0$ GeV are shown with different signs for parameters
$a$ and $b^\prime$ (full curves). It shows 
that the Pomeron exchange (dashed curve in Fig. 1(a)) plays 
dominant role, and accounts for the cross sections at forward angles. 
The resonances (dotted curves) 
and the $\pi^0$ exchange (dot-dashed curve in Fig. 1(a))
 only give small contributions to the cross sections. 
The heavy-dotted curve in Fig. 1(a) is the cross sections 
for $E_\gamma=6.45$ GeV, where the resonance effects and $\pi^0$ exchange
are so small that their contributions are negligible. Therefore, 
the pure Pomeron exchange accounts for the data (diamond), through which 
the parameter $\beta_0=1.27$ for the Pomeron-constituent-quark coupling 
is determined. With the same $\beta_0$, the Pomeron exchange 
terms are applied to the low energy region, i.e. $E_\gamma=2.0$ GeV. 

In the case of the cross section, since the non-diffractive 
contributions are so small in comparison of the dominant Pomeron 
exchange, we do not expect that clear manifestation can be derived 
for the resonance effects. However, we can see below, that the 
effects from the small cross sections can be amplified in polarization 
observables that makes it possible to investigate the role played 
by the non-diffractive resonance contributions. 

The beam polarization asymmetry  $ \check{\Sigma}$ at 
$E_\gamma$ = 2.0~GeV is shown in Fig. 2. 
Comparing the Pomeron exchange (dashed curve in Fig. 2(a))
with the Pomeron plus $\pi^0$ exchange 
(dotted curve in Fig. 2(a)), 
we find that the contribution from $\pi^0$ exchange is negligible.  
The {\it s}- and {\it u}-channel contributions 
amplified by the Pomeron exchange,
due to the interference terms, increase the magnitude
of the asymmetries by about a factor of 3 around 110$^\circ$
and produces a sign change 
above 150$^\circ$ (dot-dashed curve in Fig. 2(a)).
The three mechanisms together produce the full curves 
in Fig. 2(a) and (b). 
We present the results for the four phase sets for parameters 
$a$ and $b^\prime$ in Fig. 2(b) for comparison.

In Fig. 3, predictions for the target polarization asymmetry 
$\check{T}\equiv {\bf P}_N\cdot\hat{y}{\mathcal T}$, due to the same mechanisms
discussed above in the case of the $ \check{\Sigma}$ observable are
reported. 
It is worthy noting that the helicity amplitude structure
of the $ \check{\Sigma}$ observable differs drastically from those
of the other single polarization observables. As summarized in the 
Appendix of Ref.~\cite{NPA99}, 
the $ \check{\Sigma}$ observable is a bilinear combination
of real-real or imaginary-imaginary parts of the helicity elements, 
while the other three
single polarization observables depend on real-imaginary couples.
As the Pomeron exchange amplitude is treated purely imaginary 
in this model, the pure Pomeron exchange leads to a zero 
asymmetry (dashed curve in Fig. 3(a)) for the target polarization observable.
The $\pi^0$ exchange is purely real.
Therefore, when the $\pi^0$ exchange is added, the interference between 
the Pomeron exchange and the $\pi^0$ exchange 
produces non-zero effects (dotted curve in Fig. 3(a)). 
The Pomeron plus resonances contributions 
(dot-dashed curve in Fig. 3(a)) gives even a larger
asymmetry in magnitude than the Pomeron plus $\pi^0$ exchange does.
The full calculation (full curves in Fig. 3(a) and (b))
shows a minimum around 20$^\circ$ due to $\pi^0$ exchange and
a maximum around 130$^\circ$ generated by the resonance terms.
In both cases the Pomeron exchange plays an amplifying role in
the predicted asymmetries. 
It shows that the target 
polarization asymmetry is governed mainly by the resonance contributions 
at large angles. For comparison, we also present the results 
with phase changes in Fig. 3(b). 

In the target polarization observable, 
we find that a large cancellation arises 
between the longitudinal and transverse parts of the asymmetry, 
which produces 
a nearly zero asymmetry at $\sim 65^\circ$. This structure 
is independent on the relative phase between the Pomeron exchange 
and the $\pi^0$ exchange amplitudes, since the Pomeron exchange 
amplitude is purely imaginary and the $\pi^0$ exchange is purely 
real, therefore, the phase change will only give an overall sign to the 
dotted curve in Fig. 3(a).

The vector meson polarization and the recoil polarization 
observables are shown in Fig. 4 and Fig. 5, respectively.

In the single polarization observables, the Pomeron exchange 
mechanism turns out to be an efficient amplifier for the 
non-diffractive mechanisms suppressed in the cross sections.
Although the influence of the  $\pi^0$ exchange 
can be amplified in some polarization observables, 
it plays in general a rather minor role. 
Therefore, this might imply that the double-counting from duality
(if it exits) is negligible.
The nodal structure of the observables depends (in some cases strongly)
on the signs of the two couplings $a$ and $b^\prime$.
At large angles, it is the interferences from the 
 {\it s}- and {\it u}-channel resonances that produce significant
asymmetries.

Given the availability of polarized beam and polarized target,
we now concentrate on the beam-target (BT) double polarization
asymmetry. Another motivation in investigating this observable
is that a recently developed strangeness knock-out model~\cite{Titov} 
suggests that a small $s\overline{s}$ component ($\sim 5\%$) in the
proton might result in large asymmetries ($\sim 25$-$45\%$) 
in the BT observable at small angles. 
However, since the 
resonance contributions have not been taken into account there,
an interesting question is: if contributions from the {\it s}- 
and {\it u}-channel can produce a significant double polarization
asymmetry without introducing 
strangeness component or not.

Our predictions are shown in Fig. 6.
The Pomeron exchange alone (dashed curve in (a)), gives
a small negative asymmetry at forward angles. But 
at large angles, the asymmetry goes to about $-0.4$.
The inclusion of $\pi^0$ exchange (dotted curve in (a)) 
does not change the asymmetry significantly. However, we find that
the resonances contributions have quite strong interference
with the Pomeron exchange terms (dot-dashed curve
in (a)). The full calculation leads finally to a 
decreasing behavior, going from almost zero at 
forward angles to $ \sim -0.7$ at 180$^\circ$. This result 
(full curve in Fig. 6(a) and (b)) is obtained
with $a=-0.15$ and $b^\prime=+0.3$. The backward angle effects are
also large in the case of $a=+0.15$ and 
$b^\prime=-0.3$ (dot-dashed curve in Fig. 6(b)).
The situation becomes very different for the  
couplings sets with the same signs: 
the effect is suppressed for $a=+0.15$ and $b^\prime=+0.3$ 
(dotted curve in Fig. 6(b)),
and the shape changes drastically for $a=-0.15$ and $b^\prime=-0.3$
(dashed curve in Fig. 6(b)).
The latter set produces (almost) vanishing values at extremely backward angles.
However, the common feature of the four sets is that 
at forward angles, only small BT asymmetries are produced by the 
{\it s}- and {\it u}-channel resonance excitations though
such an interference produces dramatic changes at large angles.
This result makes it interesting 
that at forward angles, large asymmetries in the BT observable
might be able to provide some hints for 
other non-diffractive sources, especially, the possible strangeness
contents in nucleons.

%
% ----- Conclusions -----
%
\section{ Discussion and Conclusion}

In the above sections, the calculations of the nucleonic resonance
effects have been done 
in the $SU(6)\otimes O(3)$ symmetry limit. 
However, at the energy of the $\phi$ meson production threshold, 
the NRCQM symmetry is not a good approximation any more, and
large configuration mixings are expected. 
Thus, one question which must be answered in the intestigation of 
the role of the intermediate resonances is that, ``what are the effects 
produced by the configuration mixings?" Meanwhile, concerning 
the Pomeron exchange terms, which plays a role of amplifying, 
another question that one has to answer is, ``what are the effects 
from different Pomeron exchange amplitudes due to different gauge fixing 
schemes?" 
These two aspects have been taken into account in our 
submitted work, and we refer the readers to 
Ref.~\cite{NPA99} for detailed discussions. 
Here, we just summarize the main points because of the limited page space.

\begin{enumerate}
\item{\bf Pomeron gauge invariance effects}:

In the Pomeron exchange model, one meets the problem of fixing the 
gauge for the quark loop tenser. Therefore, several schemes are 
introduced. 
With the pure Pomeron exchange, these schemes 
are  consistent with each other at small angles. However, 
they behave differently at large angles. 
Therefore, the polarization asymmetries, which come from the interferences
between the diffractive Pomeron exchange terms and the non-diffractive 
amplitudes, might depend on the gauge fixing schemes, and make the 
predictions trivial, especially at large angles. 
However, with the non-diffractive contributions taken into account, we find
that the polarization observables investigated
 are not sensitive to the Pomeron structures,  
i.e. the asymmetries exhibit consistent behaviors even when different 
gauge fixing schemes are employed, and have little dependence
on the Pomeron exchange model not only at small angles, but also 
at large angles. 

\item{\bf Configuration mixing effects}:

The present data for $\gamma p\to \phi p$ near threshold are too scarce 
to allow a study of possible deviations from the $SU(6)\otimes O(3)$ symmetry.
As we discussed in Ref.~\cite{NPA99}, a reasonable alternative 
is to derive the configuration mixing coefficients for $\gamma p\to \phi p$
with an analogy to $\gamma p\to\omega p$ of which recent data 
from SAPHIR~\cite{SAPHIR}
can provide a first glance at the possible configuration mixings.
In $\gamma p\to \omega p$, for each resonance, $C_R$ is introduced for the 
amplitude, i.e.  
$h^J_{a\lambda_V}\to C_R \ h^J_{a\lambda_V} $, where
$C_R\ne 1$ means the deviation from the $SU(6)\otimes O(3)$ symmetry.
Then, the derived $C_R$ is used in $\gamma p\to \phi p$ to investigate the 
configuration mixing effects in the polarization observables.  
We find that the configuration mixings only produce small asymmetries 
at small angles though at large angles, the mixings can change 
the asymmetries significantly.

\end{enumerate}

In summary, in the $\phi$ meson 
photoproduction near threshold, 
we find that some polarization observables are not sensitive to 
resonance conrtibutions at forward angles though the asymmetries 
from the resonance interferences become significant at large angles. 
Therefore, 
at forward angles, the insensitivities of polarization observables 
to the {\it s}- and {\it u}-channel non-diffractive process imply that 
observations at small angles might be able to pin down the asymmetries 
from other channels, e.g. small strangeness component in nucleons.
At large angles, the significant sensitivities might provide 
insight to the non-diffractive $\phi$ meson production mechanism.

\noindent{\large\bf Figure captions}

\begin{itemize}

\item Fig. 1. 
Differential cross sections for $\gamma p\to \phi p$.
Data at $E_\gamma$ = 2.0~GeV (full circle) come from Ref.~\cite{phidata},
and at $E_\gamma$ = 6.45~GeV (diamond) from Ref.~\cite{phi1978}.
The heavy dotted curve in (a) is the Pomeron exchange at $E_\gamma$ = 6.45~GeV,
while all the other curves are produced at $E_\gamma$ = 2.0~GeV.
The curves are: i) $\pi^0$-exchange 
(dot-dashed); ii) {\it s}- and {\it u}-channel contribution (dotted);
iii) Pomeron exchange(dashed), and iv) contributions from i) to iii) 
(full curves). 

\item Fig. 2.
The polarized beam asymmetry at $E_\gamma$ =2.0~GeV
with different phase signs. The curves in (a) stand for:
Pomeron exchange (dashed),
Pomeron and $\pi^0$ exchanges (dotted),
Pomeron exchange and resonance contributions (dot-dashed),
and the full calculation including all three components
with $a=-0.15$, $b^\prime=0.3$ (full).
In (b), the results of the full calculations for the four
($a$, $b^\prime$) sets are depicted.

\item Fig. 3.
Same as Fig. 2, but for the polarized target asymmetry. 

\item Fig. 4.
Same as Fig. 2, but for the polarized vector meson asymmetry. 

\item Fig. 5. 
Same as Fig. 2, but for the recoil polarization asymmetry. 

\item Fig. 6.
Same as Fig. 2, but for the beam-target double polarization asymmetry.

\end{itemize}

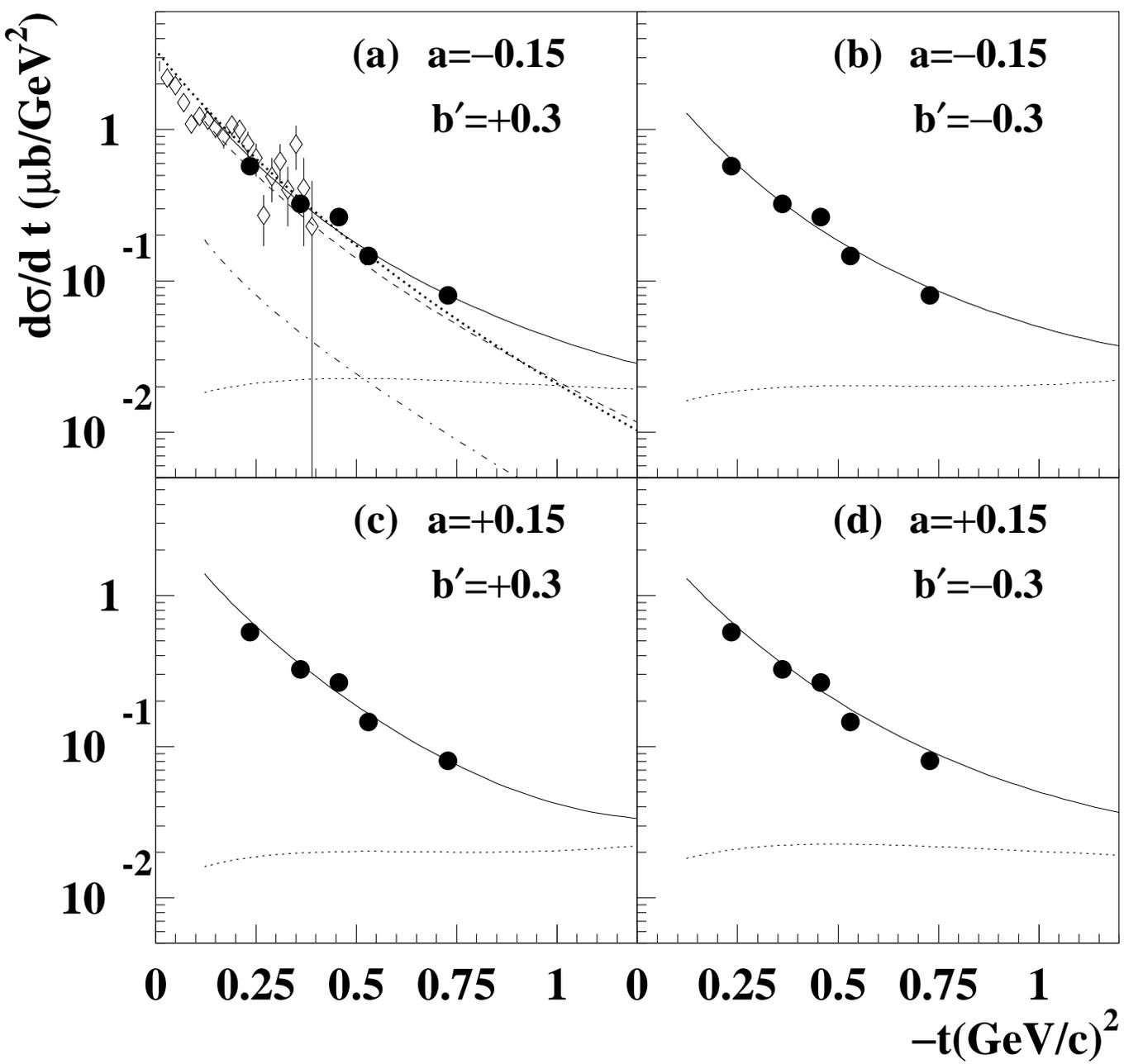
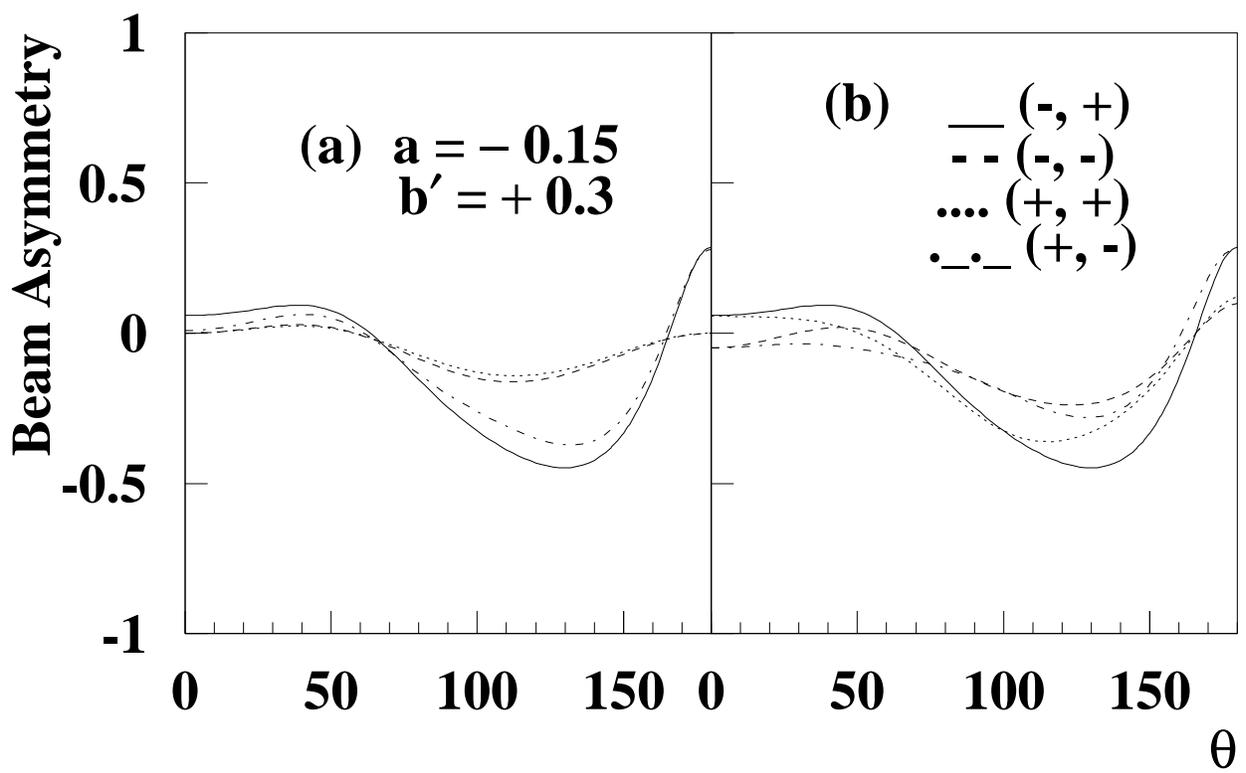
\end{document}